\documentclass[
    ,final            % use final for the camera ready runs
  ]
  {aipproc}

\layoutstyle{6x9}

\begin{document}

\title{$R$-process Nucleosynthesis during the Magnetohydrodynamics Explosions of a Massive Star}


\author{Motoaki Saruwatari}{
  address={Department of Physics, School of Sciences, Kyushu University, Fukuoka
812-8581, Japan.}
}

\author{Masa-aki Hashimoto}{
  address={Department of Physics, School of Sciences, Kyushu University, Fukuoka
812-8581, Japan.}
}

\author{Kei Kotake}{
  address={Division of Theoretical Astronomy, National Astronomical Observatory of Japan, 2-21-1 Osawa, Mitaka, Tokyo 181-8588, Japan}
}
\author{Shoichi Yamada}{
  address={Science \& Engineering, Waseda University, 3-4-1 Okubo, Shinjyuku,
Tokyo, 169-8555, Japan.}
   % additional visiting address
}

\begin{abstract}
We investigate the possibility of the r-process during the magnetohydrohynamical explosion of supernova in a massive star of 13 solar mass with the effects of neutrinos induced. We adopt five kinds of initial models which include properties of rotation and the toroidal component of the magnetic field . The simulations which succeed the explosions are limitted to a concentrated magnetic field and strong differential rotation. Low $Y_{e}$ ejecta produce heavy elements and the third peak can be reprocuced. However, the second peak is low because $Y_{e}$ distribution as a function of radius is steep and ejecta corresponding to middle $Y_{e}$ is very few.
\end{abstract}

\maketitle

%%%%%%%%%%%%%%%%%%%%%%%%%%%%%%%%%%%%%%%%%%%%
%% MAINMATTER
%%%%%%%%%%%%%%%%%%%%%%%%%%%%%%%%%%%%%%%%%%%%

\section{Introduction}
\begin{table}[b]
\vspace{-5mm}
\begin{tabular}{lrrrrrrr}
\hline
  & \tablehead{1}{r}{b}{$T/|W|$~(\%)}
  & \tablehead{1}{r}{b}{$E_m/|W|$~(\%)}
  & \tablehead{1}{r}{b}{$X_0 (10^8 \rm{cm})$}
  & \tablehead{1}{r}{b}{$Z_0 (10^8 \rm{cm})$}    
  & \tablehead{1}{r}{b}{$\Omega_0~(\rm s^{-1})$}
  & \tablehead{1}{r}{b}{$B_0$~(G)} \\
\hline
model 1 & 0   & 0    &  0   & 0 & 0   &   0 \\
model 2 & 0.5 & 0.1  &  1   & 1 & 5.2 & 5.4 $\times 10^{12}$\\
model 3 & 0.5 & 0.1  &  0.5 & 1 & 7.9 & 1.0 $\times 10^{13}$\\
model 4 & 0.5 & 0.1  &  0.1 & 1 & 42.9 & 5.2 $\times 10^{13}$\\
model 5 & 1.5 & 0.1  &  0.1 & 1 & 72.9 & 5.2 $\times 10^{13}$\\
\hline
\end{tabular}
\caption{Initial parameters of precollapse models.}
\label{tab:ini}
\end{table}
It has been considered that the origin of heavy neutron-rich elements like uranium is mainly due to the $r$-process
nucleosynthesis that occurs during the supernova explosions and/or neutron star
mergers \cite{see,sato,th01,qi03}. 
The main issue concerning the $r$-process research is to reproduce
the three peaks ($A\simeq$ 80, 130, and 195) in the abundance pattern for the $r$-elements
in the solar system.
Among models of the $r$-process, it has been believed that supernovae are the most
plausible astrophysical site \cite{qi05}.

In the present paper, we will carry out the calculations of the MHD explosion of the He-core of 3.3 $M_{\odot}$ during $t_f\sim500$~ms whose mass in the main sequence stage is about 13 $M_{\odot}$ star. For the MHD calculations, five distributions of initial rotation and magnetic fields are assumed with parametric forms. We include the effects of neutrinos by using Leakage scheme \cite{rip81}. 
Thereafter, we investigate the r-process nucleosynthesis using the
results of MHD calculations. 
%%%%%%%%%%%
%\section{SUPERNOVA MODELS}
\section{Initial Models}
The presupernova model has been calculated from the evolution of He-core of 3.3 $M_\odot$ that
corresponds to 13 $M_\odot$ in the main sequence stage \cite{hashi1995}. 
We adopt cylindrical properties of the angular velocity $\Omega$ and the toroidal component of the magnetic field $B_{\phi}$ as follows \cite{ko04a}:

\begin{equation}
	\Omega(X,Z)=\Omega_0\times {X_0^2 \over X^2+X_0^2}\cdot {Z_0^4 \over Z^4+Z_0^4}, \ \ \
	B_\phi(X,Z)=B_0\times {X_0^2 \over X^2+X_0^2}\cdot {Z_0^4 \over Z^4+Z_0^4}
\end{equation}
where $X$ and $Z$ are the distances from the rotational axis and the equatorial plane with
$X_0$ and $Z_0$ being model parameters.

Both $\Omega_0$ and $B_0$ are the initial values at $X=0$ and $Z=0$.
Initial parameters of five precollapse models are given in Table~\ref{tab:ini}. 
The spherically symmetric case is denoted by model 1.
In model 2, the profiles of rotation and magnetic field in the Fe-core 
are taken to be nearly uniform.
We present model 4 and 5 as the case having a differentially rapid rotating core and strong magnetic fields. An intermediate example, model 3 between model 2 and 
model 4 is prepared for reference. 

\section{Explosion Models and $r-$process nucleosynthesis}

We perform the calculations of the collapse, bounce, and the propagation of the shock wave with use of ZEUS-2D \cite{sn92} in which the
realistic equation of state \cite{shen98} has been implemented by Kotake et al\cite{ko04a}.
It is noted that the contribution of the nuclear energy generation
is usually negligible compared to the shock energy.
In Table~\ref{tab:res}, our results of MHD calculations are summarized.

The simulations which succeed the explosions are model 4 and model 5. Explosion fails for spherical (model1) and/or large $X_{0}$ models, which indicate that explosions need concentrated magnetic field and strong differential rotation.
Model 4 eject large area around rotation axis, but this area has high $Y_{\rm e}$. Theafore elements responsible for the third peak  do not appear.
On the other hand, Model 5 ejects collimated jet and this ejecta from deep area has very low $Y_{\rm e}$. 
Therefore we investigate the possibility of the r-process for model 5.
Low $Y_{\rm e}$ ejecta produces heavy elements and the third peak can be reprocuced(Fig.~\ref{abund}).
However, the second peak is low because $Y_{\rm e}$ distribution as a function of radius is steep and ejecta corresponding to middle $Y_{\rm e}$ is very few(Fig.~\ref{yetm}).
  
\begin{table}
\vspace{-10mm}
\caption{Hydrodynamics simulation results.}

\begin{tabular}{lrrrrrrr}
\hline
  & \tablehead{1}{r}{b}{$t_b$}
  & \tablehead{1}{r}{b}{$T/|W|_{f}$}
  & \tablehead{1}{r}{b}{$E_m/|W|_{f}$}
  & \tablehead{1}{r}{b}{$ E^*_{\rm{exp}}$}
  & \tablehead{1}{r}{b}{$M_{\rm{ej}}/M_{\odot}$}
  & \tablehead{1}{r}{b}{ $M_{\rm{rej}}/M_{\odot}$}\\
\hline
model 1 & 111 & 283  &     0   &     0      &  0.023   &   -   & \\
model 2 & 125 & 311  &  6.91   & 0.053      &  0.127   &   -   & \\
model 3 & 129 & 329  &  8.74   & 0.116      &  0.164   &   -   & \\
model 4 & 133 & 433  &  8.80   & 0.142      &  1.13    & 0.111 & \\
model 5 & 180 & 548  &  15.3   & 0.339      &  0.484   & 0.022 & \\\hline\\

\end{tabular}
\vspace{-5mm}
\footnotetext{$t_b$ indicates the time (ms) at the bounce. The calculations are stopped at the time $t_f$~(ms). The ratios $T/|W|_{f}$ and $E_m/|W|_{f}$ are expressed in \%.
$E^*_{\rm exp}=E_{\rm exp}/10^{51}$~ergs.}
\label{tab:res}
\end{table}
\begin{ltxfigure}
\vspace{-5mm}
\begin{minipage}{7.1cm} %{0.4\hsize}
%\begin{center}
\includegraphics[width=7cm,keepaspectratio]{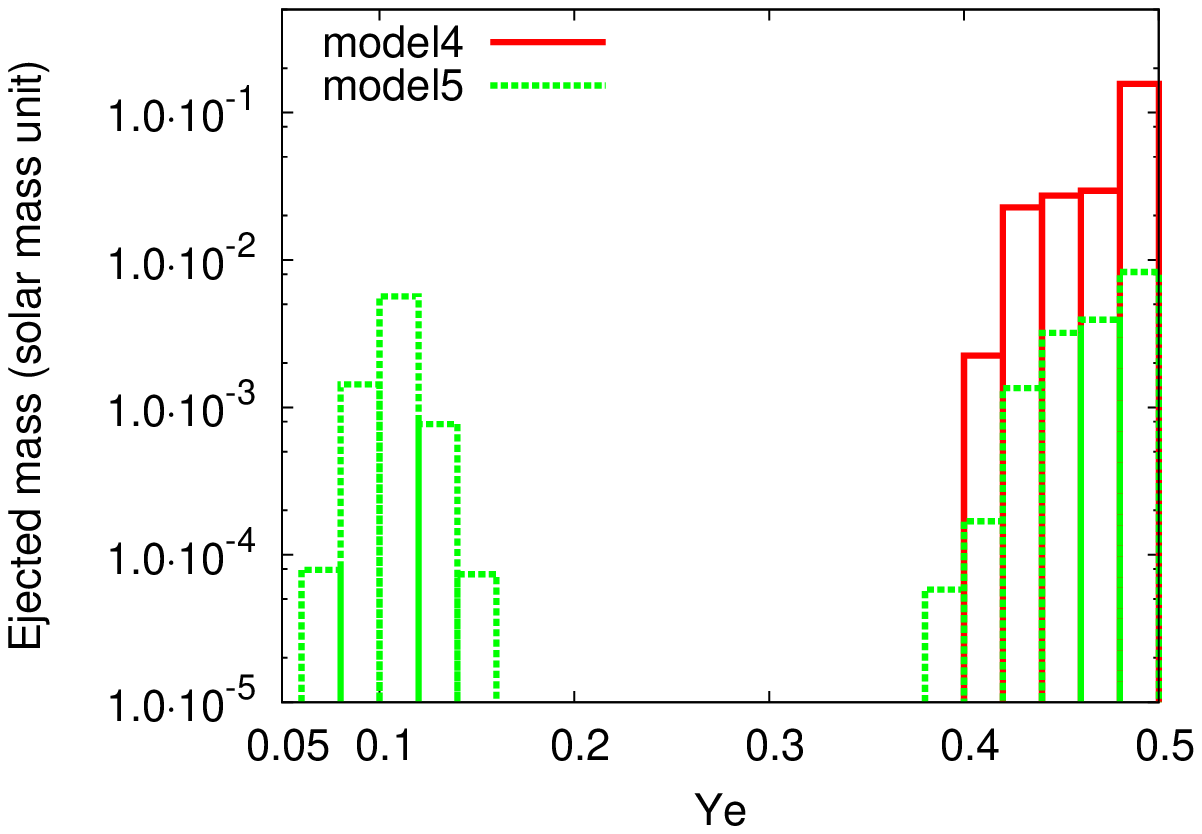}
%\end{center}
%\end{center}
\vspace{-2mm}
\caption{Ejected mass as a function of $Y_{\rm e}$ (model4 and model5).}
\label{yetm}
\end{minipage}
\begin{minipage}{7.1cm} %{0.4\hsize}
\vspace{-3mm}
%\hspace{-10mm}
%\begin{center}
\vspace{-3mm}
\includegraphics[width=7cm,keepaspectratio]{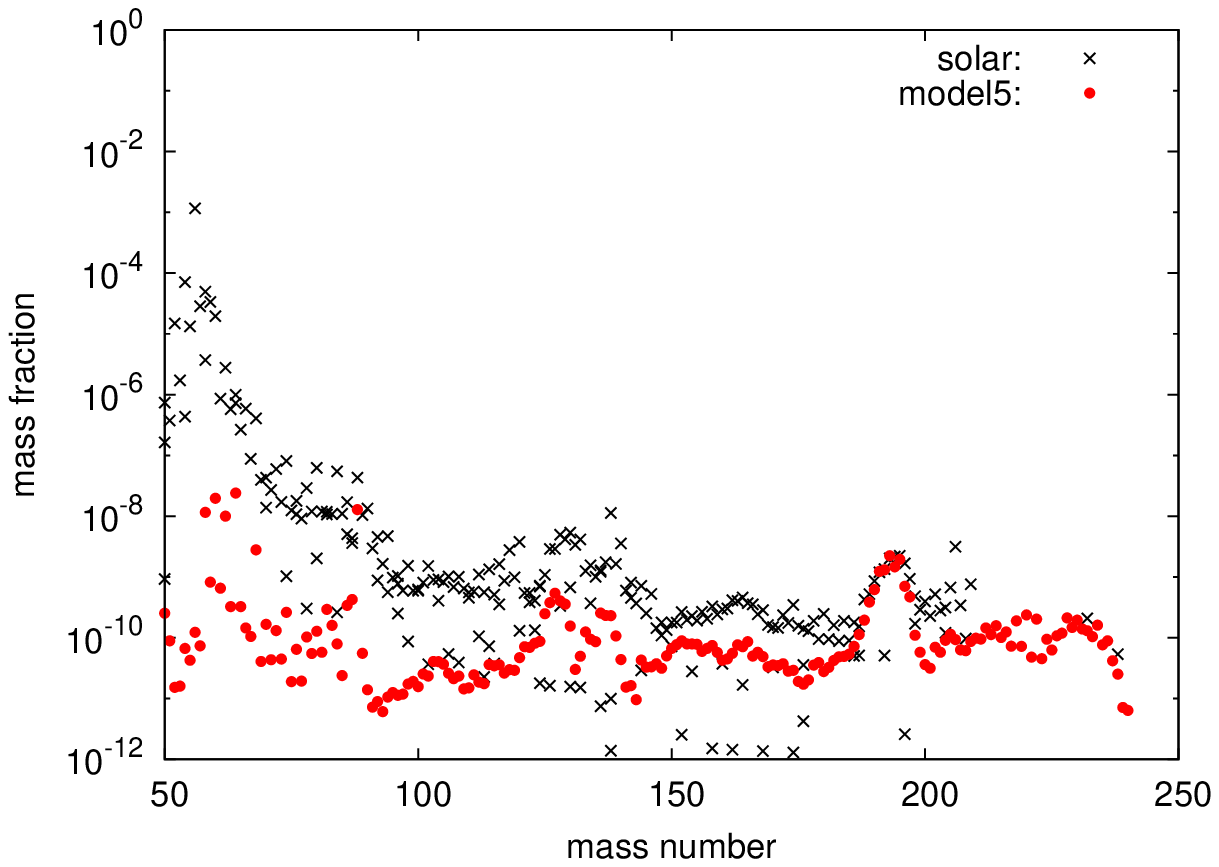}
%\end{center}
\vspace{-2mm}
\caption{Abundances obtained from model5.}
\label{abund}
\end{minipage}
\vspace{-5mm}
\end{ltxfigure}
\vspace{-2mm}
\section{SUMMARY}
\vspace{-3mm}
We calculated magneto-hydrodynamical simulations with neutrino effects included and
investigated the $r$-process nucleosynthesis
during the purely magnetohydrodynamic explosion 
in a massive star of 13 $M_{\odot}$ progenitor model by \cite{hashi1995}.
Weak differential rotation model does not succeed in explosion. So, explosion need very strong and concentrated rotation, and magnetic field.
Low $Y_{\rm e}$ ejecta from which the third peak elements appear needs strong differential rotation and concentrated magnetic field.
Explosion depend on the parameter of magnetic field and rotation profile and $Y_{\rm e}$ distribution of ejecta is very different, which depend on each parameter.
For model 5, third peak can be reprocuced but the second peak is low because $Y_{\rm e}$ distribution as a function of radius is steep.

\end{document}